\documentclass{article}
\usepackage{ijcai25}

\usepackage{times}
\usepackage{soul}
\usepackage{url}
\usepackage[hidelinks]{hyperref}
\usepackage[utf8]{inputenc}
\usepackage[small]{caption}
\usepackage{graphicx}
\usepackage{amsmath}
\usepackage{amsthm}
\usepackage{booktabs}
\usepackage{algorithm}
\usepackage{algorithmic}
\usepackage[switch]{lineno}

\usepackage{cleveref}
\Crefname{figure}{Figure}{Figures}
\crefname{figure}{Figure}{Figures}
\Crefname{table}{Table}{Tables}
\crefname{table}{Table}{Tables}
\Crefname{equation}{Equation}{Equations}
\crefname{equation}{Equation}{Equations}
\usepackage{makecell}
\usepackage{multirow}
\usepackage{tabularx}

\title{ChatMol: A Versatile Molecule Designer Based on the Numerically Enhanced Large Language Model}

\author{
Chuanliu Fan
\and
Ziqiang Cao\and
Zicheng Ma\and
Nan Yu\and
Yimin Peng\and\\
Jun Zhang\and
Yiqin Gao\And
Guohong Fu\\
\affiliations
\emails
20234027004@stu.suda.edu.cn
}

\begin{document}

\maketitle

\begin{abstract}
Goal-oriented \emph{de novo} molecule design, namely generating molecules with specific property or substructure constraints, is a crucial yet challenging task in drug discovery.
Existing methods, such as Bayesian optimization and reinforcement learning, often require training multiple property predictors and struggle to incorporate substructure constraints.
Inspired by the success of Large Language Models (LLMs) in text generation, we propose ChatMol, a novel approach that leverages LLMs for molecule design across diverse constraint settings.
Initially, we crafted a molecule representation compatible with LLMs and validated its efficacy across multiple online LLMs.
Afterwards, we developed specific prompts geared towards diverse constrained molecule generation tasks to further fine-tune current LLMs while integrating feedback learning derived from property prediction.
Finally, to address the limitations of LLMs in numerical recognition, we referred to the position encoding method and incorporated additional encoding for numerical values within the prompt.
Experimental results across single-property, substructure-property, and multi-property constrained tasks demonstrate that ChatMol consistently outperforms state-of-the-art baselines, including VAE and RL-based methods.
Notably, in multi-objective binding affinity maximization task, ChatMol achieves a significantly lower $\mathrm{K_D}$ value of $0.25$ for the protein target ESR1, while maintaining the highest overall performance, surpassing previous methods by $4.76\%$.
Meanwhile, with numerical enhancement, the Pearson correlation coefficient between the instructed property values and those of the generated molecules increased by up to $0.49$.
These findings highlight the potential of LLMs as a versatile framework for molecule generation, offering a promising alternative to traditional latent space and RL-based approaches.\footnote{The code, data, and pre-trained model will be publicly available in the final version.}
\end{abstract}

\section{Introduction}
Exploring the chemical space of small molecules to discover new drugs and materials is a pivotal issue in pharmacology and AI-assisted science research~\cite{sanchez2018inverse}.
The primary challenge in the field of pharmacology is goal-oriented \emph{de novo} drug molecule design~\cite{du2022molgensurvey}, namely generating novel and diverse drug molecules with specific biochemical properties. 
Typical examples involve designing compounds based on target features $y$, which can represent a single property, multiple properties, or a combination of substructure and property characteristics, such as optimizing molecules for high binding affinity to a designated protein target~\cite{segler2018generating}.

In light of this challenge, various AI-based methods have been developed with frameworks of variational autoencoders (VAEs)~\cite{kingma2013auto}, sequence-based language models (LMs)~\cite{rumelhart1986learning,hochreiter1997long}, and generative adversarial networks (GANs)~\cite{goodfellow2014generative}. 
Meanwhile, various optimization methods are used to generate molecules towards desirable properties, including Bayesian optimization~\cite{eckmann2022limo,kong2023molecule}, reinforcement learning\cite{you2018graph,luo2021graphdf}, and genetic algorithms~\cite{jensen2019graph}.
However, most existing approaches rely on training additional property predictors to incorporate conditions indirectly~\cite{eckmann2022limo} and struggle with conditional generation involving specific substructures~\cite{jin2018junction,you2018graph,kong2023molecule}.
Considering the inherent difficulty in accurately representing the combination of multiple property predictors~\cite{you2018graph}, coupled with the high sensitivity of molecule properties to even minor structural changes~\cite{kipf2016semi}, it is crucial to develop a solution that effectively tackles both challenges simultaneously.

\begin{figure*}[ht]
\centering
\includegraphics[width=0.85\textwidth]{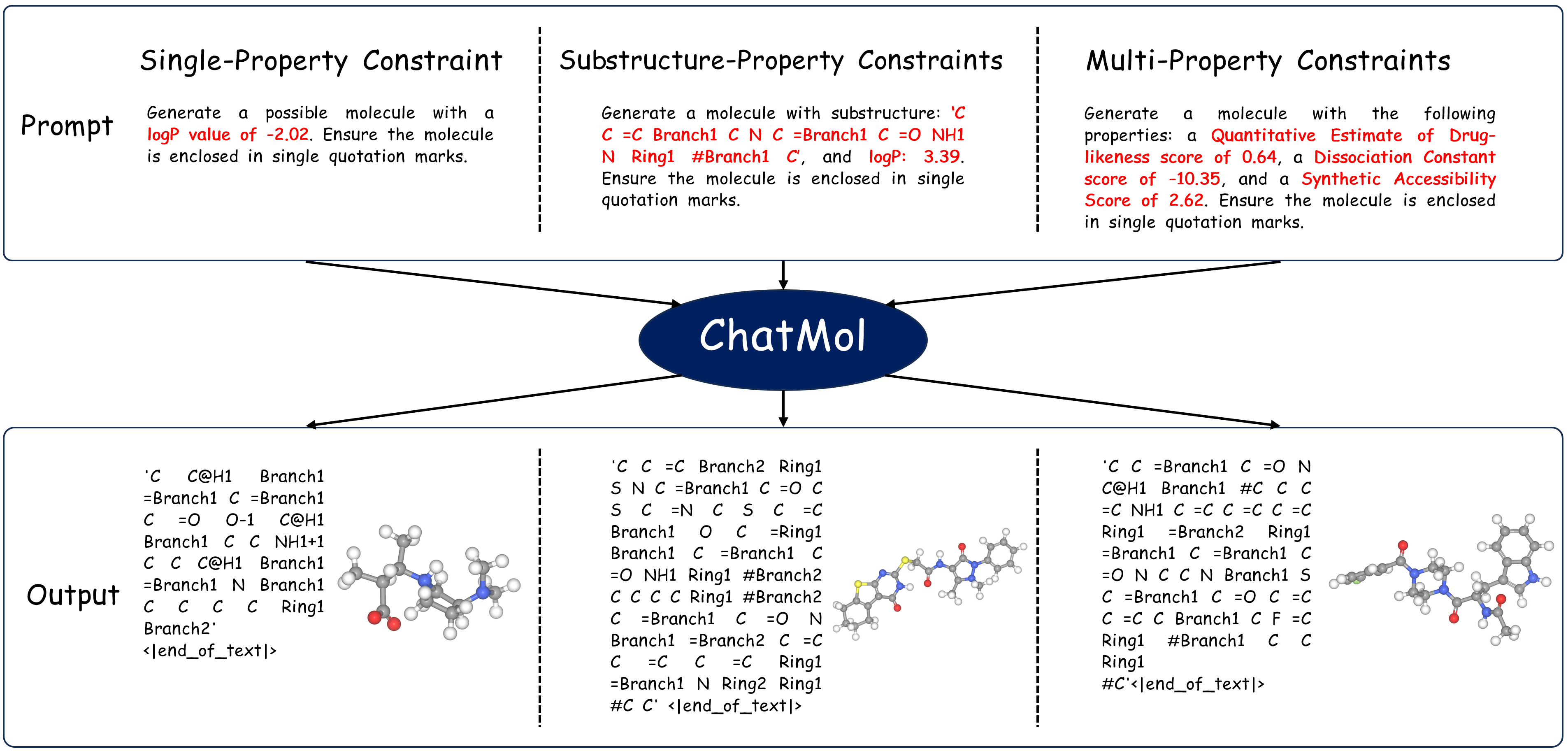}
\caption{Examples of diverse conditional generation tasks in drug design addressed by ChatMol: single-objective logP targeting, substructure-constrained logP optimization, and multi-objective binding affinity maximization.}
\label{fig_prompt}
\end{figure*}

Large Language Models (LLMs) have been successfully applied to molecule-caption translation tasks~\cite{edwards2022translation,pei2023biot5,li2024empowering}, including molecule understanding and text-based molecule generation.
For example, MolReGPT uses in-context few-shot learning to leverage the generalization capabilities of LLMs like ChatGPT~\cite{openai2024gpt4technicalreport} without requiring domain-specific pre-training or fine-tuning.
However, our preliminary experiments revealed that even SOTA online LLMs struggled to replicate incomplete molecules in drug discovery tasks, as shown in~\cref{tab_representation}.
This highlights two key issues.
First, existing molecule representations are suboptimal for LLMs. 
SMILES lacks sufficient robustness, while SELFIES generates excessively long and complicated sequences, making both unsuitable for efficient model processing.
Second, LLMs currently lack the ability to fully understand the complex conditions required in drug design, especially when dealing with numerical constraints in drug design tasks.
These limitations hinder the application of LLMs in drug discovery.

Given the problems above, we propose ChatMol, a versatile molecule designer that optimizes an LLM for \emph{de novo} molecule design.
Firstly, we simplified SELFIES to make the molecule representation more concise and closer to natural language while preserving its inherent high validity.
Experiments conducted on three online LLMs revealed that the proposed molecule representation offers significant advantages in inference efficiency and achieves a high success rate according to~\cref{tab_representation}.
Secondly, we developed specific prompts geared towards diverse constrained molecule generation tasks to further fine-tune current LLMs while integrating feedback learning derived from property prediction. 
Finally, inspired by the position encoding method~\cite{vaswani2017attention}, we applied a unified numerical encoding to numbers within the prompt, enhancing the model's ability to follow the numerical instructions.

We tested LLM backbones of varying sizes, ranging from 1B to 8B parameters~\cite{dubey2024llama3herdmodels}, to evaluate the scalability of our approach.
Experimental results across single-property, substructure-property, and multi-property constrained tasks verfied the effectiveness of ChatMol.
In the multi-objective binding affinity maximization task, ChatMol achieves a significantly lower $\mathrm{K_D}$ value of $0.25$, while maintaining the highest overall performance, surpassing previous methods by $4.76\%$.
Furthermore, our ablation study demonstrated that numerical enhancement significantly improves the ability of the model to follow numerical instructions, increasing the Pearson correlation coefficient by up to $0.49$.
Notably, by setting prompts according to the distribution of training set properties, our approach also achieved promising results in the goal-free de novo generation task, as detailed in Appendix A.

Our contributions are as follows: 
\begin{itemize}
    \item We proposed an end-to-end versatile molecule design framework using LLMs, capable of performing goal-oriented \emph{de novo} molecule generation tasks under various biochemical constraints.
    \item We proposed a simplified molecule representation particularly tailored for LLMs.
    \item We designed a numerical encoding method for the numbers in the conditions, which improves the alignment between the instructed property values and those of the generated drug molecules.
\end{itemize}

\section{Related Work}
\label{sec_related_work}
\subsection{Molecule Generation}
One important task in molecular design is goal-oriented \emph{de novo} molecule generation~\cite{zunger2018inverse}, which aims to design novel compounds with specific biochemical properties. 
Our approach falls within this category, focusing on the generation of molecules under predefined constraints through a non-iterative process. 
In contrast, methods like LIMO~\cite{eckmann2022limo} adopt an iterative approach to achieve similar goals.
Another key task is goal-oriented molecule optimization, exemplified by methods such as MolMIM~\cite{reidenbach2022improving}, which iteratively refines molecules starting from a reference compounds. MOLGEN~\cite{fang2023domain}, on the other hand, is a non-iterative approach that also relies on a reference molecule as the starting point for optimization.
Finally, unconditional de novo molecule generation~\cite{flam2022language,jin2018junction,liu2019hyperbolic} involves the generation of novel molecules without specific property constraints, where the model generates molecules that conform to the statistical distribution of the training data.
A detailed comparison of these generative tasks is presented in \cref{tab_generative_tasks}.

\begin{table*}[ht] 
\centering 
\normalsize
\begin{tabular}{lccc}
\toprule 
Generative Task & Reference Molecule & Model & Optimization Method \\
\midrule 
\multirow{2}{*}{Goal-oriented \emph{de novo} generation} & w/o & LIMO & Iterative \\
& w/o & \textbf{Ours} & Non-Iterative \\
\midrule
\multirow{2}{*}{Goal-oriented optimization} & w/ & MolMIM & Iterative \\
& w/ & MOLGEN & Non-Iterative \\
\midrule
Goal-free \emph{de novo} generation & w/o & SF-RNN & - \\
\bottomrule 
\end{tabular} 
\caption{Comparison of different generative tasks in molecule design. ``w/'' indicates that the input includes a reference molecule as the starting point, while ``w/o'' means that no reference molecule is provided in the input. Goal-free generation involves no constraints on molecule properties or substructures, whereas goal-oriented generation incorporates specific constraints.}
\label{tab_generative_tasks} 
\end{table*}

\begin{figure*}[ht]
\centering
\includegraphics[width=0.85\textwidth]{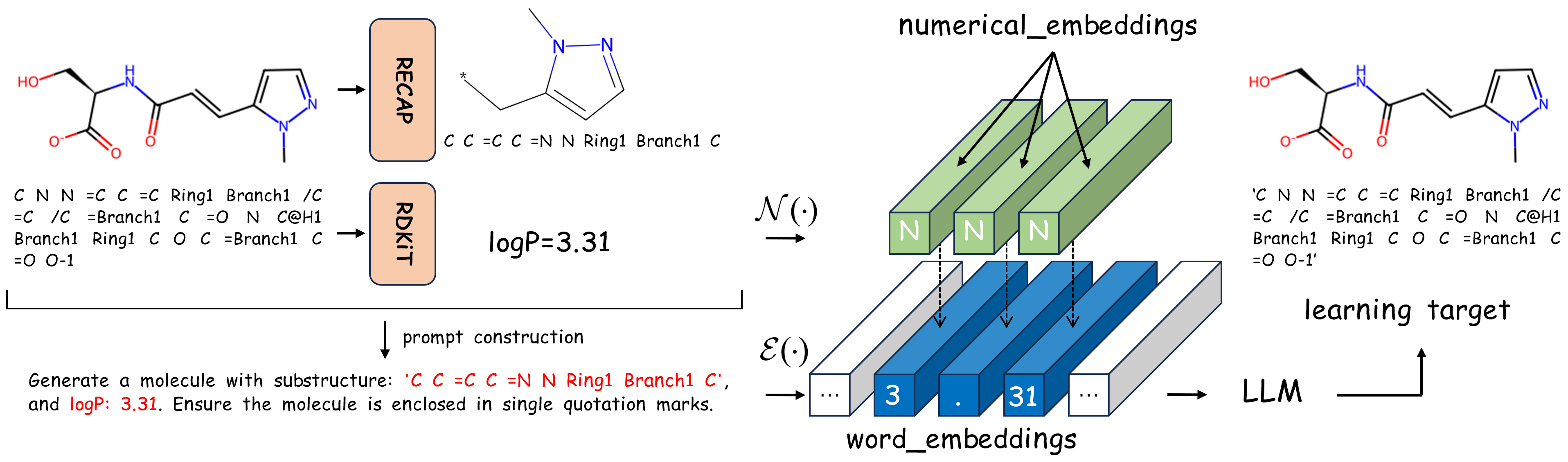}
\caption{Illustration of numerical enhancement in the training process. Property values are transformed through $\mathcal{N}(\cdot)$ to obtain a holistic numerical encoding, which is then added to each numerical token's word embedding to produce the final encoding of the constraint conditions. $\mathcal{E}(\cdot)$ represents the embedding layer.}
\label{fig_dataset_preprocess}
\end{figure*}

\subsection{Molecule Representations}
From the perspective of generation objectives, inverse molecule design includes 3D molecule generation~\cite{hoogeboom2022equivariant,bao2022equivariant}, 2D graph-based molecule generation~\cite{mahmood2021masked,seff2019discrete}, and 1D sequence-based molecule generation~\cite{flam2022language,segler2018generating}.
There are several inverse molecule design methods focus on 3D molecule generation, including conditional autoregressive model~\cite{gebauer2022inverse}, diffusion models~\cite{bao2022equivariant,hoogeboom2022equivariant}. 
However, we focus on 1D string-based representation that uses simplified text encodings to describe the structure of a chemical species, which is highly interpretable compared to 3D coordinates. 
Sequence representation also offers computational efficiency and facilitates the exploration of longer and more diverse molecules~\cite{fang2023domain}.

\section{Method}
\label{sec_method}
ChatMol generates molecules end-to-end by taking in a text description of desired properties and substructures and producing molecules in a string-based representation. 
To achieve this goal, we first designed a molecule representation tailored for LLMs, focusing on robustness and simplicity. 
Then, we improved numerical representation by adding numerical embeddings to the token embeddings for different numbers.
The model was trained using a two-stage process.

\subsection{Molecule Representation}
\label{subsec_represent}
SMILES and SELFIES are the most commonly used string-based representations of a molecule~\cite{weininger1988smiles,krenn2020self}.
SMILES possesses the advantage of conciseness, while it lacks a mechanism to ensure the validity of molecule strings in terms of syntax and physical principles. 
Conversely, the square brackets in SELFIES representation, without expanding the vocabulary of LLM, often result in more tokens and conflicts during tokenization. 
To ensure both conciseness and chemical validity, we propose a simplified SELFIES representation by removing the square brackets and using spaces to separate each element, thus achieving a format closer to natural language for molecule representation. 
As illustrated in~\cref{tab_tokenization}, the tokenization results for Llama3 8B on the ZINC250k~\cite{irwin2012zinc} and MOSES~\cite{10.3389/fphar.2020.565644} dataset indicate that our approach produces an average token count half that of SELFIES representation.
Consequently, we improved the molecule representation without expanding the vocabulary, significantly lowering the barrier to use.

\begin{table}[ht] 
\centering 
\normalsize
\begin{tabular}{lccc} 
\toprule 
Representation & SMILES & SELFIES & \textbf{Ours} \\
\midrule 
$\bar{L}_{\text{ZINC250k}}$ & 29.34 & 99.45 & 61.99 \\
$\bar{L}_{\text{MOSES}}$ & 24.27 & 89.67 & 55.00 \\
\bottomrule 
\end{tabular} 
\normalsize 
\caption{Comparison of average tokenization lengths for different molecular representations on the ZINC250k and MOSES datasets.}
\label{tab_tokenization} 
\end{table}

Furthermore, We tested three online LLMs using different types of molecule representations on the single-property task, i.e., generating molecules with logP targeted within the range of ($-2.5, -2$).
For each inference, we provided three examples, conducting a total of $212$ In-Context Learning inferences. 
Notably, the inference times across different LLMs are not directly comparable due to potential differences in inference configurations and network speeds. 
However, within each model, the comparison of inference times for the three representations highlights the effectiveness of our proposed method. 

As illustrated in \cref{tab_representation}, our representation demonstrated the highest effectiveness in terms of success rates.
Moreover, it was observed that the proposed representation consistently exhibited less inference times compared to SELFIES, and even surpassed SMILES on ERNIE.
These ICL inference results, obtained without fine-tuning, further confirmed that the proposed simplified SELFIES representation is more suitable for Large Language Models, both in terms of effectiveness and efficiency.

\begin{table}[ht]
\centering
\small
\begin{tabular}{llccc}
\toprule
Metric & Model & SMILES & SELFIES & \textbf{Ours} \\
\midrule
\multirow{5}{*}{Success(\%)} & ERNIE  & $0.47\%$ & $0.47\%$ & $\mathbf{3.77\%}$ \\
      & moonshot  & $11.32\%$ & $33.96\%$ & $\mathbf{38.68\%}$ \\
      & GPT-4o & $7.81\%$ & $7.69\%$ & $\mathbf{11.63\%}$ \\
      & MolT5 & $3.93\%$ & - & - \\
      & MolReGPT & $1.00\%$ & - & - \\
\midrule
\multirow{5}{*}{Time(s)} & ERNIE  & 418.18 & 584.63 & \textbf{396.52} \\
      & moonshot  & \textbf{538.39} & 4080.42 & 2280.56 \\
      & GPT-4o & \textbf{243.21} & 385.21 & 348.77 \\
      & MolT5 & 1212.12 & - & - \\
      & MolReGPT & 702.93 & - & - \\
\bottomrule
\end{tabular}
\normalsize
\caption{In-Context Learning inference performance of MolT5~\protect\cite{edwards2022translation}, MolReGPT~\protect\cite{li2024empowering} and online LLMs (ERNIE-Lite-8K-0922~\protect\cite{sun2020ernie}, moonshot-v1-8k~\protect\cite{qin2024mooncake}, and GPT-4o~\protect\cite{openai2024gpt4technicalreport}) with different molecule representations on the task of single-objective logP targeting.}
\label{tab_representation}
\end{table}

\subsection{Numerical Enhancement}
\label{sub_ne}
We perform numerical enhancement for each number that appears in the input conditions. 
Inspired by the absolute positional encoding method~\cite{vaswani2017attention}, we use the following formula to obtain the numerical embedding, which is then added to the corresponding part of the text embedding. 
As shown in \cref{eq_numerical_enhancement}, assume that the number $num$ is tokenized into three tokens, and the dimension of the embedding hidden layer is $d_{hidden\_size}$. 
Then we have:
\begin{equation}
\begin{aligned}
\mathcal{N}{(num, 2i)} &= \sin{(num/10000^{2i/{d_{hidden\_size}}})} \\
\mathcal{N}{(num, 2i+1)} &= \cos{(num/10000^{2i/{d_{hidden\_size}}})}
\end{aligned}
\label{eq_numerical_enhancement}
\end{equation}
Here, $\mathcal{N}$ represents the Numerical Embedding, $num$ is the numerical value, and $i$ ranges from $0$ to $1/2 \times d_{hidden\_size}$. The odd dimensions of the hidden layer use sine encoding, while the even dimensions use cosine encoding.

After deriving the numerical embedding based on the number, we add it to the embeddings of all tokens corresponding to that number in the condition prompt. Suppose $\mathcal{T}_{num}$ is the set of numerical tokens, then the updated embedding for each token $j \in \mathcal{T}_{num}$ is given by:
\begin{equation}
\mathbf{E}_{token}^{j} = \mathbf{E}_{token}^{j} + \mathcal{N}, \quad \forall j \in \mathcal{T}_{num}
\label{eq_numerical_enhancement2}
\end{equation}
Where $\mathbf{E}_{token}^{j}$ denotes the token embedding of the number $num$.
Since we have replaced the position with the number in the conventional positional encoding, all the tokens of the number will share the same numerical embedding, thereby enhancing the numerical information as a whole.

\subsection{Two-Stage Training}
\label{sub_sft}
The two-stage training process is illustrated in Appendix B.
The goal of neural molecule design is to create a function $g$ parameterized by a causal language model with parameter $\theta$ that takes a condition $C$ consisting of property or substructure constraints and generates an appropriate molecule $M$ described above. 
Denote $\{C, M^*\}$ is a specific training sample, tokens of the reference molecule $M^{*}$ is $\{t^{*}_1,...,t^{*}_l\}$ with a length $l$ after tokenization. The training objective is to minimize the sum of the negative log-likelihoods associated with these tokens. The cross-entropy loss can be expressed as:

\begin{align}
\mathcal{L}_{ce}=-\sum_{j=1}^{l} \sum_t p_{true} \left(t \mid C, M_{<j}^{*}\right) \log p_{g_{\theta}}\left(t \mid C, M_{<j}^{*};\theta\right)
\label{eq_ce_loss}
\end{align}
where $t$ is the possible next token from the vocabulary, and $M_{<j}^{*}$ is the partial reference token sequence $\{t^*_1,...,t^*_{j-1}\}$.
While $p_{g_{\theta}}\left(t \mid C, M_{<j}^{*};\theta\right)$ is the probability the model $g_\theta$ generates the corresponding token $t$. We sum all the probability of possible token $t$. 
And $p_{\text{true}}\left(t \mid C, M_{<j}^{*}\right) = 1$ if $t = t_{j}^{*}$, and $p_{\text{true}}\left(t \mid C, M_{<j}^{*}\right) = 0$ if $t \neq t_{j}^{*}$, where $t_j^*$ is the reference next token of the partial reference molecule $M_{<j}^{*}$ given condition $C$.

Following \cite{liu2022brio}, we require ChatMol to possess the capability to precisely predict the ranking order of a set of the sampled candidate molecules $\hat{\mathbf{M}}$, derived from the model trained in the supervised fine-tuning stage.
The fundamental principle of sequence calibration involves generating multiple candidate results $\hat{M} \in \hat{\mathbf{M}}$ initially, followed by ranking these candidates using a custom discriminator $S$.
After we rank the candidate molecules based on the $S(\hat{M})$, we can fine-tune the model from the SFT stage with a ranking loss~\cite{hopkins2011tuning,zhong2020extractive} as follows:
\begin{align} 
\mathcal{L}_{rank}=\sum_{i} \sum_{j>i} max\left(0, f_{\theta}(\hat{M}_j) - f_{\theta}(\hat{M}_i)\right)
\label{eq_ranking_loss}
\end{align}
where $\hat{M}_i$ and $\hat{M}_j$ are two candidate molecules from $\hat{\mathbf{M}}$ and satisfy $\forall i,j,i<j$, $S(\hat{M}_i) > S(\hat{M}_j)$. $f_{\theta}(\hat{M})$ denotes the estimated log-probability of the candidate molecule $\hat{M}$ provided by ChatMol with parameter $\theta$.
To implement \ref{eq_ranking_loss},
we first generate six candidate molecules for each condition input of the training sample considering the constraints of cost, and then score them.

The quality of candidate molecules can be assessed through the measurement scorer $S$, which comprises the property scorer $S_p$ and the structure scorer $S_s$. 
ChatMol is capable of handling both property and substructure constraints, as well as their combinations. Taking single property and substructure as an example, when given the target property $y^*$, we first calculate the property of each candidate molecule $\hat{M} \in \hat{\mathbf{M}}$ using RDKit~\cite{landrum2006rdkit} or AutoDock-GPU~\cite{santos2021accelerating}, denoted by $f_{\alpha}(\hat{M})$.

The property score can be defined as $S_p(\hat{M}) = 1-|y^* - f_{\alpha}(\hat{M})|/max(y^*, f_{\alpha}(\hat{M}))$. Given the desired substructure $M_{frag}^*$, the structure score $S_s(\hat{M}) = 1$ if the candidate molecule contains the substructure, and $S_s(\hat{M}) = 0$ otherwise. We use RDKit to determine whether candidate molecules contain the target substructure. Then we calculate the weighted scores $S(\hat{M})= w_p \cdot S_p(\hat{M}) + w_s \cdot S_s(\hat{M})$.

To mitigate the influence of molecule length on the score $f_{\theta}(\hat{M})$, we normalize the log-probability based on the molecule's length, following the approach proposed by \cite{cho2014properties}:
\begin{align}
f_{\theta}(\hat{M})=\frac{\sum_{j=1}^{l} \log p_{g_{\theta}} \left(\hat{t}_j \mid \hat{C}, \hat{M}_{<j} ; \theta \right)}{|\hat{M}|^{\beta}}
\end{align}
where $\hat{C}$ is the condition of the six candidate molecules and $\hat{t}$ is the next token of the partial candidate molecule $\hat{M}_{<j}$. The length penalty hyperparameter $\beta$ is set to $2.0$ following \cite{liu2022brio}.
 
The model fine-tuned solely with ranking loss may no longer be used as a generative model~\cite{liu2022brio}, so we merged the sequence-level ranking loss with token-level cross-entropy loss to preserve the generation ability of ChatMol:
\begin{align}
\mathcal{L}_{total}=\gamma_{1}\mathcal{L}_{ce} + \gamma_{2}\mathcal{L}_{rank}
\end{align}
following the setting in \cite{cho2014properties}, we set $\gamma_{1} = 0.1$ and $\gamma_{2} = 10.0$ due to the difference in their magnitudes.

\section{Experiments}
\label{sec_exp}
\subsection{Dataset}
For all molecule generation tasks, we use ZINC250k as our dataset, which contains $\approx 250,000$ purchasable, drug-like molecules.
We utilized AutoDock-GPU to generate docking scores for a SMILES string against two protein targets: the human estrogen receptor (ESR1) and human peroxisomal acetyl-CoA acyltransferase 1 (ACAA1).
We employed the RECAP (Retrosynthetic Combinatorial Analysis Procedure) algorithm~\cite{lewell1998recap} to fragment molecules in the ZINC250k dataset for substructure-related training, which is a computational technique designed to electronically fragment molecules based on chemical knowledge. Molecules lacking leaf nodes were excluded.
Finally, we convert the SMILES sequences into the simplified SELFIES format as described in the molecule representation subsection.
The size of the training sets for each task is as follows:
\begin{itemize}
    \item Single-objective logP targeting: 636 samples.
    \item Substructure-property constrained logP extremization: 250k samples.
    \item Multi-objective binding affinity maximization: 1,987 samples (ESR1) and 828 samples (ACAA1).
\end{itemize}

\subsection{Tasks and Metrics}
In the single-objective logP targeting task, we aimed to generate molecules with logP in the range $(-2.5, -2)$, associated with favorable pharmacokinetics~\cite{eckmann2022limo}. We defined two metrics: success rate (proportion of molecules within the desired logP range) and diversity (one minus the average pairwise Tanimoto similarity of Morgan fingerprints~\cite{rogers2010extended}).

In the substructure-property constrained logP extremization task, the goal is to generate molecules containing a specified substructure with extreme logP properties.
The evaluation metrics verify whether the generated molecules contain the specified substructures and meet the logP value requirements.

The multi-objective binding affinity maximization task focuses on optimizing multiple drug properties, including targeting, pharmacokinetics, and synthesis feasibility.
We defined an overall metric as $\log(\frac{QED}{K_D \times SA})$, integrating QED (higher scores for better drug suitability), $\mathrm{K_D}$ (lower values for better binding affinity~\cite{santos2021accelerating}), and the SA score (lower values for easier synthesis~\cite{bickerton2012quantifying,ertl2009estimation}).

\subsection{Experimental Setup}
The prompts for the three tasks are set as shown in \cref{fig_prompt}. We determine the prompts based on the range of properties required for each task. The prompts contain numerous numbers, which motivates the proposal of numerical enhancement.

The metrics for downstream tasks of drug discovery primarily focus on the maximum or minimum value of a specific property or require the property to fall within a certain range, which cannot fully reflect the model's adherence to prompts. We will discuss the impact of the numerical enhancement in the last subsection.

\subsection{Baselines}
We compare with the following works for conditional molecule design, VAE-based Bayesian optimization method JT-VAE~\cite{jin2018junction}, and LIMO~\cite{eckmann2022limo}, reinforcement learning-based method GCPN~\cite{you2018graph}, and LSTM-based Bayesian optimization method SGDS~\cite{kong2023molecule}. 
LIMO is the only approach capable of handling both properties and substructures.

\subsection{Single-Objective logP Targeting}
\label{sub_logp_targeting}
In this task, our prompt includes a logP value randomly sampled between $-2.5$ and $-2$ with two decimal places.
As shown in \cref{tab_logp_targeting}, ChatMol outperforms other methods significantly in terms of success rate, achieving an impressive $94.5\%$, though it exhibits a slightly lower diversity. The success rates of models based on both types of backbones have improved after sequence calibration, with models utilizing larger parameters holding a greater advantage.

\begin{table}[ht]
\centering
\small
\begin{tabular}{lcc}
\toprule
\textbf{Method} & \textbf{Success ($\%$)} & \textbf{Diversity} \\
\midrule
ZINC & 0.4 & \textbf{0.919} \\
\midrule
JT-VAE & 11.3 & 0.846 \\
GCPN & 85.5 & 0.392 \\
LIMO & 10.4 & 0.914 \\
SGDS & 86.0 & 0.874 \\
\midrule
\textbf{ChatMol} (1B) & 57.8 & 0.704 \\
\textbf{ChatMol} $\dagger$ & 81.4 & 0.701 \\
\textbf{ChatMol} & \textbf{94.5} & 0.703 \\
\bottomrule
\end{tabular}
\normalsize
\caption{Results for single-objective property targeting to $-2.5 <$ logP $< -2.0$. ZINC is the distribution of the ZINC250k dataset and results for JT-VAE, GCPN, LIMO, and SGDS are taken from~\protect\cite{kong2023molecule}. $\dagger$ denotes without feedback training.}
\label{tab_logp_targeting}
\end{table}

\subsection{Substructure-Property Constrained logP Extremization}
\label{sub_structure}
In this task, our inference prompt includes a substructure sequence and a property value.
Following \cite{eckmann2022limo}, we selected the same two starting molecules with the specified substructures identified by them.
We sampled the property values around the extreme logP values of the training set distribution, which is far away from the logP values of the two starting molecules.

\begin{figure}[ht]
\centering
\includegraphics[width=\linewidth]{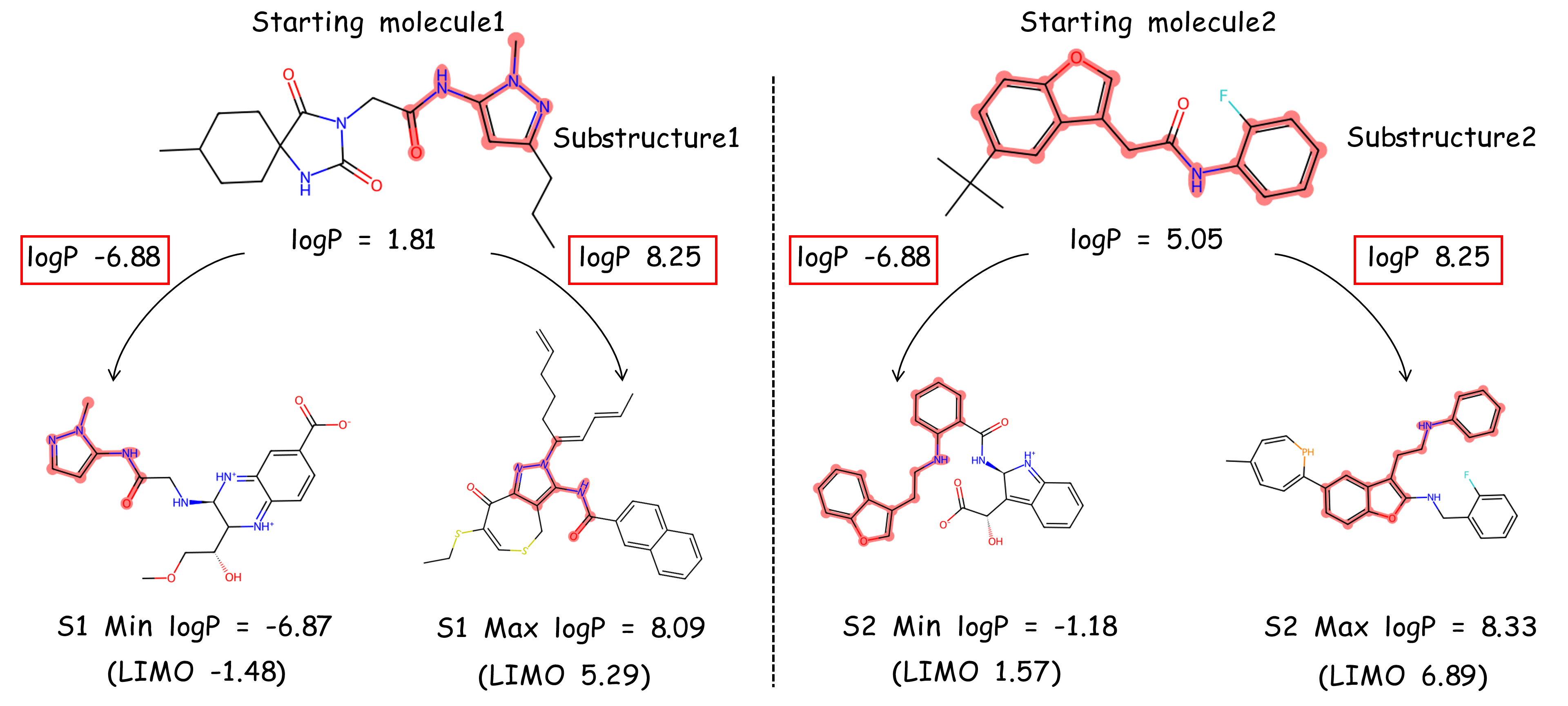}
\caption{Extremization of logP property with substructure (highlighted in red) fixed. The logP value within the red box represents our target value, and together with the substructure, it constitutes the prompt for this task.}
\label{fig_substructure}
\end{figure}

\cref{fig_substructure} illustrates the logP extremization results, the starting logP values are 1.81 and 5.05 respectively.
As shown in~\cref{tab_prop_substructure}, compared to LIMO, ChatMol expanded the range of logP values, while maintaining the substructures of the two starting molecules fixed. Although KIMI (moonshot-v1-8k) can generate some molecules through ICL, the properties of the generated molecules show a significant weak correlation with the target properties. Moreover, KIMI fails when encountering complex long molecular structures like substructure 2.

We also calculated the Pearson correlation coefficient $\rho$ between the logP values set in the prompt and the actual logP values of the generated molecules within the whole inference batch. As shown in~\cref{tab_prop_substructure}, the strong correlations indicate that our model effectively adheres to the desired property values in this task.

\begin{table}[ht]
\centering
\setlength{\tabcolsep}{1mm}
\small
\begin{tabular}{lcccccc}
\toprule
\textbf{Method} & \textbf{S1 Min} & \textbf{S1 Max} & \textbf{S1} $\rho$ & \textbf{S2 Min} & \textbf{S2 Max} & \textbf{S2} $\rho$ \\
\midrule
LIMO & -1.48 & 5.29 & - & 1.57 & 6.89 & - \\
KIMI & -2.39 & \textbf{11.14} & 0.01 & - & - & - \\
\midrule
\textbf{ChatMol} (1B) & -4.35 & 7.62 & 0.83 & 0.95 & 6.46 & 0.20 \\
\textbf{ChatMol} $\dagger$ & -5.13 & 7.74 & 0.85 & 1.04 & 6.28 & 0.63 \\
\textbf{ChatMol} & \textbf{-6.87} & 8.09 & \textbf{0.93} & \textbf{-1.18} & \textbf{8.33} & \textbf{0.78} \\
\bottomrule
\end{tabular}
\normalsize
\caption{Comparison of logP extremization and Pearson correlation coefficient $\rho$ under the substructure-property constraints, S1 and S2 denote Substructure1 and Substructure2 specified by LIMO~\protect\cite{eckmann2022limo}. $\dagger$ denotes without feedback training.}
\label{tab_prop_substructure}
\end{table}

Although our model was trained on fragments from the RECAP algorithm and had not encountered these two specific substructures before, it is still capable of generating molecules containing these novel substructures.
This highlights our model's strong generalization ability and confirms the effectiveness of incorporating structural information through our training approach.

\subsection{Multi-Objective Binding Affinity Maximization}
\label{sub_multi}
In this task, our inference prompt includes three property values, each sampled near the extreme values in their respective training set distributions, depending on the directions of the target properties.
Following \cite{eckmann2022limo}, we target the binding sites of two human proteins: estrogen receptor (PDB ESR1, UniProt P03372) and peroxisomal acetyl-CoA acyl transferase 1 (PDB ACAA1, UniProt P09110) and generate 100k molecules for each protein. Our objective is to generate molecules that exhibit high binding affinity (low $\mathrm{K_D}$ values), a QED score greater than $0.4$, and a SA score below $5.5$. We generate 100k molecules for each protein target and present the top-2 scores for $\mathrm{K_D}$, QED, and SA.

\begin{table*}[ht]
\centering
\small
\begin{tabular}{lcccccccc}
\toprule
\multirow{2}{*}[-0.3ex]{\textbf{Ligand}} & \multicolumn{4}{c}{\textbf{ESR1}} & \multicolumn{4}{c}{\textbf{ACAA1}} \\
 & $\mathrm{K_D} (\downarrow)$ & QED $(\uparrow)$ & SA $(\downarrow)$ & overall $(\uparrow)$ & $\mathrm{K_D} (\downarrow)$ & QED $(\uparrow)$ & SA $(\downarrow)$ & overall $(\uparrow)$ \\
\midrule
Tamoxifen & 87 & 0.45 & \textbf{2.0} & 14.77 & - & - & - & - \\
Raloxifene & $7.9\times10^6$ & 0.32 & \underline{2.4} & 2.83 & - & - & - & - \\
\midrule
GCPN $1^{st}$ & 810 & 0.43 & 4.2 & 11.75 & 8500 & \underline{0.69} & 4.2 & 9.87 \\
GCPN $2^{nd}$ & $2.7\times10^4$ & \textbf{0.80} & 3.7 & 8.99 & 8500 & 0.54 & 4.3 & 9.60 \\
LIMO $1^{st}$ & 4.6 & 0.43 & 4.8 & 16.78 & 28 & 0.57 & 5.5 & 15.12 \\
LIMO $2^{nd}$ & 2.8 & 0.64 & 4.9 & 17.66 & 31 & 0.44 & 4.9 & 14.88 \\
SGDS $1^{st}$ & \underline{0.36} & 0.44 & 3.99 & 19.54 & 4.55 & 0.56 & 4.07 & 17.22 \\
SGDS $2^{nd}$ & 1.28 & 0.44 & 3.86 & 18.30 & 5.67 & 0.60 & 4.58 & 16.96 \\
\midrule
\textbf{ChatMol} (1B) $1^{st}$& 0.69 & 0.61 & 3.85 & 19.25 & 16 & 0.62 & \underline{3.88} & 16.09 \\
\textbf{ChatMol} (1B) $2^{nd}$& 1.40 & 0.64 & 3.42 & 18.71 & 17 & \underline{0.69} & \textbf{3.83} & 16.16 \\
\textbf{ChatMol} $1^{st}$& \textbf{0.25} & 0.53 & 2.74 & \textbf{20.47} & \underline{4.48} & \textbf{0.78} & 4.24 & \textbf{17.53} \\
\textbf{ChatMol} $2^{nd}$& \underline{0.36} & \underline{0.71} & 4.17 & \underline{19.97} & \textbf{2.79} & 0.57 & 5.21 & \underline{17.48} \\
\bottomrule
\end{tabular}
\normalsize
\caption{Comparison of molecules generated for protein targets ESR1 and ACAA1 in the multi-property constrained generation task. Arrow ($\uparrow$) indicates higher scores are preferable, and ($\downarrow$) indicates lower scores are desired. We present the Top-2 scores for $\mathrm{K_D}$, QED, and SA. The best results are highlighted in bold, while the second-best results are highlighted with underline. $1^{st}$ denotes the top-ranked molecule and $2^{nd}$ represents the second-best molecule from the 100k generated molecules. Define overall as $\log(\frac{QED}{K_D \times SA})$. Baseline results are obtained from~\protect\cite{eckmann2022limo,kong2023molecule}.}
\label{tab_multi_obj}
\end{table*}

~\cref{tab_multi_obj} illustrates that ChatMol demonstrates the best overall performance metrics across the two ligand generation tasks for the target proteins, achieving the lowest $\mathrm{K_D}$ values for both ESR1 and ACAA1. 
Our ranking is based on the overall metric, so it is reasonable that the enhancement in overall performance may involve some compromise in individual metrics like SA score.

\subsection{Analysis of Numerical Enhancement}
Due to the nature of the above tasks in drug design, the evaluation metrics either focus on whether a certain molecule property falls within a specific range or on identifying the molecules with the maximum or minimum values of certain properties, which all center on individual molecules and are not related to specific numbers in the prompts.

These metrics fail to reflect how well the model adheres to the property values specified in the prompt across an entire batch of inferences. Therefore, we introduced the analysis experiments of numerical enhancement, using Pearson correlation coefficients and RMSE to assess the consistency between the target properties and the actual properties of the generated molecules.

To eliminate interference from other modules, we conducted SFT training based on the Llama3 8B model solely on the last two tasks, which had more complex prompts and involved more numerical data, with and without numerical enhancement.
As shown in \cref{tab_pearson}, the logP pertains to the substructure-property constrained logP extremization task, and $\mathrm{K_D}$, QED, and SA, are derived from the multi-property constrained task. The results with numerical enhancement are on the right, while the results without it are on the left.

The RMSE of the $\mathrm{K_D}$ value in the multi-property constrained task decreased from 3.184 to 0.684, representing a reduction of approximately $78.5\%$. Correspondingly, the Pearson correlation coefficient improved from -0.103 to 0.378, changing from a negative to a positive correlation.

\begin{table}[ht]
\centering
\normalsize
\fontfamily{ptm}\selectfont
\begin{tabular}{lcc}
\toprule
\textbf{Metric} & \textbf{RMSE ($\downarrow$)} & \textbf{$\rho$ ($\uparrow$)} \\
\midrule
\textbf{logP} & 5.409 / 2.930 & 0.736 / 0.848 \\
\textbf{$\mathrm{K_D}$} & 3.184 / 0.684 & -0.103 / 0.378 \\
\textbf{QED} & 0.301 / 0.154 & 0.047 / 0.304 \\
\textbf{SA} & 1.707 / 0.806 & 0.026 / 0.514 \\
\bottomrule
\end{tabular}
\normalsize
\caption{Comparison of RMSE and Pearson correlation coefficient $\rho$ between the target properties of the prompts and the corresponding properties of the generated molecules with (right) and without (left) numerical enhancement across two tasks.}
\label{tab_pearson}
\end{table}

Although the effect of numerical enhancement for logP was less pronounced, it was significantly more evident in the multi-property setting, demonstrating that when a prompt contains a large number of numerical values (in this case, up to three values), the proposed numerical enhancement method significantly improves the model's compliance with numerical instructions.

\section{Conclusion}
We present ChatMol, a versatile molecule designer that leverages LLMs for \emph{de novo} molecule design.
Our approach first represented molecules as simplified SELFIES sequences, offering a natural language-like format that ensures chemical validity and conciseness at a low cost.
Then, We fine-tuned Llama3 with property prediction feedback to align the model distribution with desired properties and substructures.
Finally, to address the issue of weak numerical recognition in LLMs, we implemented numerical encoding for arbitrary numbers within the condition context.
Experimental results demonstrate that our method surpasses all other baseline approaches across all three tasks.
After numerical enhancement, the Pearson correlation coefficient between the instructed property values and those of the generated molecules showed a significant increase, reaching up to $0.49$.
This demonstrates that Large Language Models can excel in molecule design tasks in a more direct, versatile, and practical manner compared to latent space generative models or RL-based methods.

We conducted preliminary exploration utilizing the Llama3 8B model. 
In accordance with the scaling law~\cite{kaplan2020scaling}, 
we are highly confident that scaling up the model parameters will lead to significant breakthroughs.

\clearpage
\bibliographystyle{named}
\bibliography{ijcai25}

\begin{thebibliography}{}

\bibitem[\protect\citeauthoryear{Bao \bgroup \em et al.\egroup }{2022}]{bao2022equivariant}
Fan Bao, Min Zhao, Zhongkai Hao, Peiyao Li, Chongxuan Li, and Jun Zhu.
\newblock Equivariant energy-guided sde for inverse molecular design.
\newblock In {\em The eleventh international conference on learning representations}, 2022.

\bibitem[\protect\citeauthoryear{Bickerton \bgroup \em et al.\egroup }{2012}]{bickerton2012quantifying}
G~Richard Bickerton, Gaia~V Paolini, J{\'e}r{\'e}my Besnard, Sorel Muresan, and Andrew~L Hopkins.
\newblock Quantifying the chemical beauty of drugs.
\newblock {\em Nature chemistry}, 4(2):90--98, 2012.

\bibitem[\protect\citeauthoryear{Cho \bgroup \em et al.\egroup }{2014}]{cho2014properties}
Kyunghyun Cho, Bart Van~Merri{\"e}nboer, Dzmitry Bahdanau, and Yoshua Bengio.
\newblock On the properties of neural machine translation: Encoder-decoder approaches.
\newblock {\em arXiv preprint arXiv:1409.1259}, 2014.

\bibitem[\protect\citeauthoryear{Du \bgroup \em et al.\egroup }{2022}]{du2022molgensurvey}
Yuanqi Du, Tianfan Fu, Jimeng Sun, and Shengchao Liu.
\newblock Molgensurvey: A systematic survey in machine learning models for molecule design.
\newblock {\em arXiv preprint arXiv:2203.14500}, 2022.

\bibitem[\protect\citeauthoryear{Dubey \bgroup \em et al.\egroup }{2024}]{dubey2024llama3herdmodels}
Abhimanyu Dubey, Abhinav Jauhri, Abhinav Pandey, and et~al. Kadian, Abhishek.
\newblock The llama 3 herd of models, 2024.

\bibitem[\protect\citeauthoryear{Eckmann \bgroup \em et al.\egroup }{2022}]{eckmann2022limo}
Peter Eckmann, Kunyang Sun, Bo~Zhao, Mudong Feng, Michael~K Gilson, and Rose Yu.
\newblock Limo: Latent inceptionism for targeted molecule generation.
\newblock {\em Proceedings of machine learning research}, 162:5777, 2022.

\bibitem[\protect\citeauthoryear{Edwards \bgroup \em et al.\egroup }{2022}]{edwards2022translation}
Carl Edwards, Tuan Lai, Kevin Ros, Garrett Honke, Kyunghyun Cho, and Heng Ji.
\newblock Translation between molecules and natural language.
\newblock {\em arXiv preprint arXiv:2204.11817}, 2022.

\bibitem[\protect\citeauthoryear{Ertl and Schuffenhauer}{2009}]{ertl2009estimation}
Peter Ertl and Ansgar Schuffenhauer.
\newblock Estimation of synthetic accessibility score of drug-like molecules based on molecular complexity and fragment contributions.
\newblock {\em Journal of cheminformatics}, 1:1--11, 2009.

\bibitem[\protect\citeauthoryear{Fang \bgroup \em et al.\egroup }{2023}]{fang2023domain}
Yin Fang, Ningyu Zhang, Zhuo Chen, Lingbing Guo, Xiaohui Fan, and Huajun Chen.
\newblock Domain-agnostic molecular generation with self-feedback.
\newblock {\em arXiv preprint arXiv:2301.11259}, 2023.

\bibitem[\protect\citeauthoryear{Flam-Shepherd \bgroup \em et al.\egroup }{2022}]{flam2022language}
Daniel Flam-Shepherd, Kevin Zhu, and Al{\'a}n Aspuru-Guzik.
\newblock Language models can learn complex molecular distributions.
\newblock {\em Nature Communications}, 13(1):3293, 2022.

\bibitem[\protect\citeauthoryear{Gebauer \bgroup \em et al.\egroup }{2022}]{gebauer2022inverse}
Niklas~WA Gebauer, Michael Gastegger, Stefaan~SP Hessmann, Klaus-Robert M{\"u}ller, and Kristof~T Sch{\"u}tt.
\newblock Inverse design of 3d molecular structures with conditional generative neural networks.
\newblock {\em Nature communications}, 13(1):973, 2022.

\bibitem[\protect\citeauthoryear{Goodfellow \bgroup \em et al.\egroup }{2014}]{goodfellow2014generative}
Ian Goodfellow, Jean Pouget-Abadie, Mehdi Mirza, Bing Xu, David Warde-Farley, Sherjil Ozair, Aaron Courville, and Yoshua Bengio.
\newblock Generative adversarial nets.
\newblock {\em Advances in neural information processing systems}, 27, 2014.

\bibitem[\protect\citeauthoryear{Hochreiter and Schmidhuber}{1997}]{hochreiter1997long}
Sepp Hochreiter and J{\"u}rgen Schmidhuber.
\newblock Long short-term memory.
\newblock {\em Neural computation}, 9(8):1735--1780, 1997.

\bibitem[\protect\citeauthoryear{Hoogeboom \bgroup \em et al.\egroup }{2022}]{hoogeboom2022equivariant}
Emiel Hoogeboom, V{\i}ctor~Garcia Satorras, Cl{\'e}ment Vignac, and Max Welling.
\newblock Equivariant diffusion for molecule generation in 3d.
\newblock In {\em International conference on machine learning}, pages 8867--8887. PMLR, 2022.

\bibitem[\protect\citeauthoryear{Hopkins and May}{2011}]{hopkins2011tuning}
Mark Hopkins and Jonathan May.
\newblock Tuning as ranking.
\newblock In {\em Proceedings of the 2011 Conference on Empirical Methods in Natural Language Processing}, pages 1352--1362, 2011.

\bibitem[\protect\citeauthoryear{Irwin \bgroup \em et al.\egroup }{2012}]{irwin2012zinc}
John~J Irwin, Teague Sterling, Michael~M Mysinger, Erin~S Bolstad, and Ryan~G Coleman.
\newblock Zinc: a free tool to discover chemistry for biology.
\newblock {\em Journal of chemical information and modeling}, 52(7):1757--1768, 2012.

\bibitem[\protect\citeauthoryear{Jensen}{2019}]{jensen2019graph}
Jan~H Jensen.
\newblock A graph-based genetic algorithm and generative model/monte carlo tree search for the exploration of chemical space.
\newblock {\em Chemical science}, 10(12):3567--3572, 2019.

\bibitem[\protect\citeauthoryear{Jin \bgroup \em et al.\egroup }{2018}]{jin2018junction}
Wengong Jin, Regina Barzilay, and Tommi Jaakkola.
\newblock Junction tree variational autoencoder for molecular graph generation.
\newblock In {\em International conference on machine learning}, pages 2323--2332. PMLR, 2018.

\bibitem[\protect\citeauthoryear{Kaplan \bgroup \em et al.\egroup }{2020}]{kaplan2020scaling}
Jared Kaplan, Sam McCandlish, Tom Henighan, Tom~B Brown, Benjamin Chess, Rewon Child, Scott Gray, Alec Radford, Jeffrey Wu, and Dario Amodei.
\newblock Scaling laws for neural language models.
\newblock {\em arXiv preprint arXiv:2001.08361}, 2020.

\bibitem[\protect\citeauthoryear{Kingma and Welling}{2013}]{kingma2013auto}
Diederik~P Kingma and Max Welling.
\newblock Auto-encoding variational bayes.
\newblock {\em arXiv preprint arXiv:1312.6114}, 2013.

\bibitem[\protect\citeauthoryear{Kipf and Welling}{2016}]{kipf2016semi}
Thomas~N Kipf and Max Welling.
\newblock Semi-supervised classification with graph convolutional networks.
\newblock {\em arXiv preprint arXiv:1609.02907}, 2016.

\bibitem[\protect\citeauthoryear{Kong \bgroup \em et al.\egroup }{2023}]{kong2023molecule}
Deqian Kong, Bo~Pang, Tian Han, and Ying~Nian Wu.
\newblock Molecule design by latent space energy-based modeling and gradual distribution shifting.
\newblock In {\em Uncertainty in Artificial Intelligence}, pages 1109--1120. PMLR, 2023.

\bibitem[\protect\citeauthoryear{Krenn \bgroup \em et al.\egroup }{2020}]{krenn2020self}
Mario Krenn, Florian H{\"a}se, AkshatKumar Nigam, Pascal Friederich, and Alan Aspuru-Guzik.
\newblock Self-referencing embedded strings (selfies): A 100\% robust molecular string representation.
\newblock {\em Machine Learning: Science and Technology}, 1(4):045024, 2020.

\bibitem[\protect\citeauthoryear{Landrum and others}{2006}]{landrum2006rdkit}
Greg Landrum et~al.
\newblock Rdkit: Open-source cheminformatics, 2006.

\bibitem[\protect\citeauthoryear{Lewell \bgroup \em et al.\egroup }{1998}]{lewell1998recap}
Xiao~Qing Lewell, Duncan~B Judd, Stephen~P Watson, and Michael~M Hann.
\newblock Recap retrosynthetic combinatorial analysis procedure: a powerful new technique for identifying privileged molecular fragments with useful applications in combinatorial chemistry.
\newblock {\em Journal of chemical information and computer sciences}, 38(3):511--522, 1998.

\bibitem[\protect\citeauthoryear{Li \bgroup \em et al.\egroup }{2024}]{li2024empowering}
Jiatong Li, Yunqing Liu, Wenqi Fan, Xiao-Yong Wei, Hui Liu, Jiliang Tang, and Qing Li.
\newblock Empowering molecule discovery for molecule-caption translation with large language models: A chatgpt perspective.
\newblock {\em IEEE Transactions on Knowledge and Data Engineering}, 2024.

\bibitem[\protect\citeauthoryear{Liu \bgroup \em et al.\egroup }{2019}]{liu2019hyperbolic}
Qi~Liu, Maximilian Nickel, and Douwe Kiela.
\newblock Hyperbolic graph neural networks.
\newblock {\em Advances in neural information processing systems}, 32, 2019.

\bibitem[\protect\citeauthoryear{Liu \bgroup \em et al.\egroup }{2022}]{liu2022brio}
Yixin Liu, Pengfei Liu, Dragomir Radev, and Graham Neubig.
\newblock Brio: Bringing order to abstractive summarization.
\newblock {\em arXiv preprint arXiv:2203.16804}, 2022.

\bibitem[\protect\citeauthoryear{Luo \bgroup \em et al.\egroup }{2021}]{luo2021graphdf}
Youzhi Luo, Keqiang Yan, and Shuiwang Ji.
\newblock Graphdf: A discrete flow model for molecular graph generation.
\newblock In {\em International conference on machine learning}, pages 7192--7203. PMLR, 2021.

\bibitem[\protect\citeauthoryear{Mahmood \bgroup \em et al.\egroup }{2021}]{mahmood2021masked}
Omar Mahmood, Elman Mansimov, Richard Bonneau, and Kyunghyun Cho.
\newblock Masked graph modeling for molecule generation.
\newblock {\em Nature communications}, 12(1):3156, 2021.

\bibitem[\protect\citeauthoryear{OpenAI \bgroup \em et al.\egroup }{2024}]{openai2024gpt4technicalreport}
OpenAI, Josh Achiam, Steven Adler, and et~al. Agarwal, Sandhini.
\newblock Gpt-4 technical report, 2024.

\bibitem[\protect\citeauthoryear{Pei \bgroup \em et al.\egroup }{2023}]{pei2023biot5}
Qizhi Pei, Wei Zhang, Jinhua Zhu, Kehan Wu, Kaiyuan Gao, Lijun Wu, Yingce Xia, and Rui Yan.
\newblock Biot5: Enriching cross-modal integration in biology with chemical knowledge and natural language associations.
\newblock {\em arXiv preprint arXiv:2310.07276}, 2023.

\bibitem[\protect\citeauthoryear{Polykovskiy \bgroup \em et al.\egroup }{2020}]{10.3389/fphar.2020.565644}
Daniil Polykovskiy, Alexander Zhebrak, Benjamin Sanchez-Lengeling, Sergey Golovanov, Oktai Tatanov, Stanislav Belyaev, Rauf Kurbanov, Aleksey Artamonov, Vladimir Aladinskiy, Mark Veselov, Artur Kadurin, Simon Johansson, Hongming Chen, Sergey Nikolenko, Alan Aspuru-Guzik, and Alex Zhavoronkov.
\newblock {M}olecular {S}ets ({MOSES}): {A} {B}enchmarking {P}latform for {M}olecular {G}eneration {M}odels.
\newblock {\em Frontiers in Pharmacology}, 2020.

\bibitem[\protect\citeauthoryear{Qin \bgroup \em et al.\egroup }{2024}]{qin2024mooncake}
Ruoyu Qin, Zheming Li, Weiran He, Mingxing Zhang, Yongwei Wu, Weimin Zheng, and Xinran Xu.
\newblock Mooncake: Kimi's kvcache-centric architecture for llm serving.
\newblock {\em arXiv preprint arXiv:2407.00079}, 2024.

\bibitem[\protect\citeauthoryear{Reidenbach \bgroup \em et al.\egroup }{2022}]{reidenbach2022improving}
Danny Reidenbach, Micha Livne, Rajesh~K Ilango, Michelle Gill, and Johnny Israeli.
\newblock Improving small molecule generation using mutual information machine.
\newblock {\em arXiv preprint arXiv:2208.09016}, 2022.

\bibitem[\protect\citeauthoryear{Rogers and Hahn}{2010}]{rogers2010extended}
David Rogers and Mathew Hahn.
\newblock Extended-connectivity fingerprints.
\newblock {\em Journal of chemical information and modeling}, 50(5):742--754, 2010.

\bibitem[\protect\citeauthoryear{Rumelhart \bgroup \em et al.\egroup }{1986}]{rumelhart1986learning}
David~E Rumelhart, Geoffrey~E Hinton, and Ronald~J Williams.
\newblock Learning representations by back-propagating errors.
\newblock {\em nature}, 323(6088):533--536, 1986.

\bibitem[\protect\citeauthoryear{Sanchez-Lengeling and Aspuru-Guzik}{2018}]{sanchez2018inverse}
Benjamin Sanchez-Lengeling and Al{\'a}n Aspuru-Guzik.
\newblock Inverse molecular design using machine learning: Generative models for matter engineering.
\newblock {\em Science}, 361(6400):360--365, 2018.

\bibitem[\protect\citeauthoryear{Santos-Martins \bgroup \em et al.\egroup }{2021}]{santos2021accelerating}
Diogo Santos-Martins, Leonardo Solis-Vasquez, Andreas~F Tillack, Michel~F Sanner, Andreas Koch, and Stefano Forli.
\newblock Accelerating autodock4 with gpus and gradient-based local search.
\newblock {\em Journal of chemical theory and computation}, 17(2):1060--1073, 2021.

\bibitem[\protect\citeauthoryear{Seff \bgroup \em et al.\egroup }{2019}]{seff2019discrete}
Ari Seff, Wenda Zhou, Farhan Damani, Abigail Doyle, and Ryan~P Adams.
\newblock Discrete object generation with reversible inductive construction.
\newblock {\em Advances in neural information processing systems}, 32, 2019.

\bibitem[\protect\citeauthoryear{Segler \bgroup \em et al.\egroup }{2018}]{segler2018generating}
Marwin~HS Segler, Thierry Kogej, Christian Tyrchan, and Mark~P Waller.
\newblock Generating focused molecule libraries for drug discovery with recurrent neural networks.
\newblock {\em ACS central science}, 4(1):120--131, 2018.

\bibitem[\protect\citeauthoryear{Sun \bgroup \em et al.\egroup }{2020}]{sun2020ernie}
Yu~Sun, Shuohuan Wang, Yukun Li, Shikun Feng, Hao Tian, Hua Wu, and Haifeng Wang.
\newblock Ernie 2.0: A continual pre-training framework for language understanding.
\newblock In {\em Proceedings of the AAAI Conference on Artificial Intelligence}, volume~34, pages 8968--8975, 2020.

\bibitem[\protect\citeauthoryear{Vaswani}{2017}]{vaswani2017attention}
Ashish Vaswani.
\newblock Attention is all you need.
\newblock {\em arXiv preprint arXiv:1706.03762}, 2017.

\bibitem[\protect\citeauthoryear{Weininger}{1988}]{weininger1988smiles}
David Weininger.
\newblock Smiles, a chemical language and information system. 1. introduction to methodology and encoding rules.
\newblock {\em Journal of chemical information and computer sciences}, 28(1):31--36, 1988.

\bibitem[\protect\citeauthoryear{You \bgroup \em et al.\egroup }{2018}]{you2018graph}
Jiaxuan You, Bowen Liu, Zhitao Ying, Vijay Pande, and Jure Leskovec.
\newblock Graph convolutional policy network for goal-directed molecular graph generation.
\newblock {\em Advances in neural information processing systems}, 31, 2018.

\bibitem[\protect\citeauthoryear{Zhong \bgroup \em et al.\egroup }{2020}]{zhong2020extractive}
Ming Zhong, Pengfei Liu, Yiran Chen, Danqing Wang, Xipeng Qiu, and Xuanjing Huang.
\newblock Extractive summarization as text matching.
\newblock {\em arXiv preprint arXiv:2004.08795}, 2020.

\bibitem[\protect\citeauthoryear{Zunger}{2018}]{zunger2018inverse}
Alex Zunger.
\newblock Inverse design in search of materials with target functionalities.
\newblock {\em Nature Reviews Chemistry}, 2(4):0121, 2018.

\end{thebibliography}

\clearpage
\appendix
\section{Additional Experiments}
\subsection{Basic Metrics}
We present the evaluation metrics for the generated molecules in comparison to each training set across three tasks: validity, uniqueness and novelty. Validity refers to the percentage of generated molecules that are chemically valid. Uniqueness denotes the percentage of generated molecules that are distinct and not duplicated. Novelty is defined as the percentage of valid molecules that do not appear in the training set. As shown in~\cref{tab_basic_metrics}, our method achieves nearly $100\%$ validity and novelty. Since structural constraints require the generated molecules to contain specific substructures, there is a higher repetition rate when calculating pairwise Tanimoto similarity. This explicit restriction results in the molecule generation task with substructure constraints performing slightly lower in terms of diversity.

\begin{table}[ht]
\centering
\small
\fontfamily{ptm}\selectfont
\begin{tabular}{lcccc}
\toprule
Tasks & Valid & Unique & Novel & Time(s/mol) \\
\midrule
Single-Property & 1.00 & 0.70 & 0.99 & 0.06 \\
Structure-Property & 1.00 & 0.62 & 1.00 & 0.33 \\
Multi-Properties & 0.99 & 0.76 & 1.00 & 0.18 \\
\bottomrule
\end{tabular}
\caption{Validity, uniqueness, and novelty of the 10,000 generated molecules compared to the training sets for each task. Valid: the percentage of molecules that are chemically valid. Unique: percentage of generated molecules that are distinct. Novel: percentage of valid generated molecules not present in training set. Time denotes the inference time per molecule on each device.}
\label{tab_basic_metrics}
\end{table}

\subsection{Goal-Free Generation of Molecules}
\label{sub_unconditional}
We also conducted additional unconditional generation experiments on the MOSES benchmark. The results from the random generation of 30,000 molecules are summarized in~\cref{tab_moses_metrics}.
We assume that the property distribution in the training set follows a normal distribution. By sampling property values from this distribution, we construct inputs to perform the random generation task.
The high uniqueness and validity provides the foundation for ChatMol’s ability to generate a diverse set of novel molecules with desirable properties.

\begin{table*}
\centering
\normalsize
\fontfamily{ptm}\selectfont
\begin{tabular}{lccccc}
\toprule
Model & Valid($\uparrow$) & Unique@1k($\uparrow$) & Unique@10k($\uparrow$) & IntDiv($\uparrow$) & Novelty($\uparrow$) \\
\midrule
JT-VAE & 1.00 & 1.00 & 0.9996 & 0.855 & 0.9143 \\
GRAPHDF & 1.00 & 1.00 & 0.9972 & 0.887 & 1.00 \\
LIMO & 1.00 & 0.998 & 0.9756 & 0.907 & 1.00 \\
\textbf{Ours} & 1.00 & 1.00 & 0.9970 & 0.895 & 0.9813 \\
\bottomrule
\end{tabular}
\caption{Unconditional generation of molecules trained on MOSES benchmark. The results are calculated with the MOSES platform~\protect\cite{10.3389/fphar.2020.565644}. Valid: percentage of molecules that are chemically valid. Unique@1k: percentage of 1,000 generated molecules that are unique. Unique@10k: percentage of 10,000 generated molecules that are unique. IntDiv: one minus average pairwise similarity between molecules. Novelty: percentage of valid generated molecules not present in training set. Results are taken from~\protect\cite{eckmann2022limo}.}
\label{tab_moses_metrics}
\end{table*}

\section{Implementation Details}
\label{sec_details}
\begin{figure*}
\centering
\includegraphics[width=0.85\textwidth]{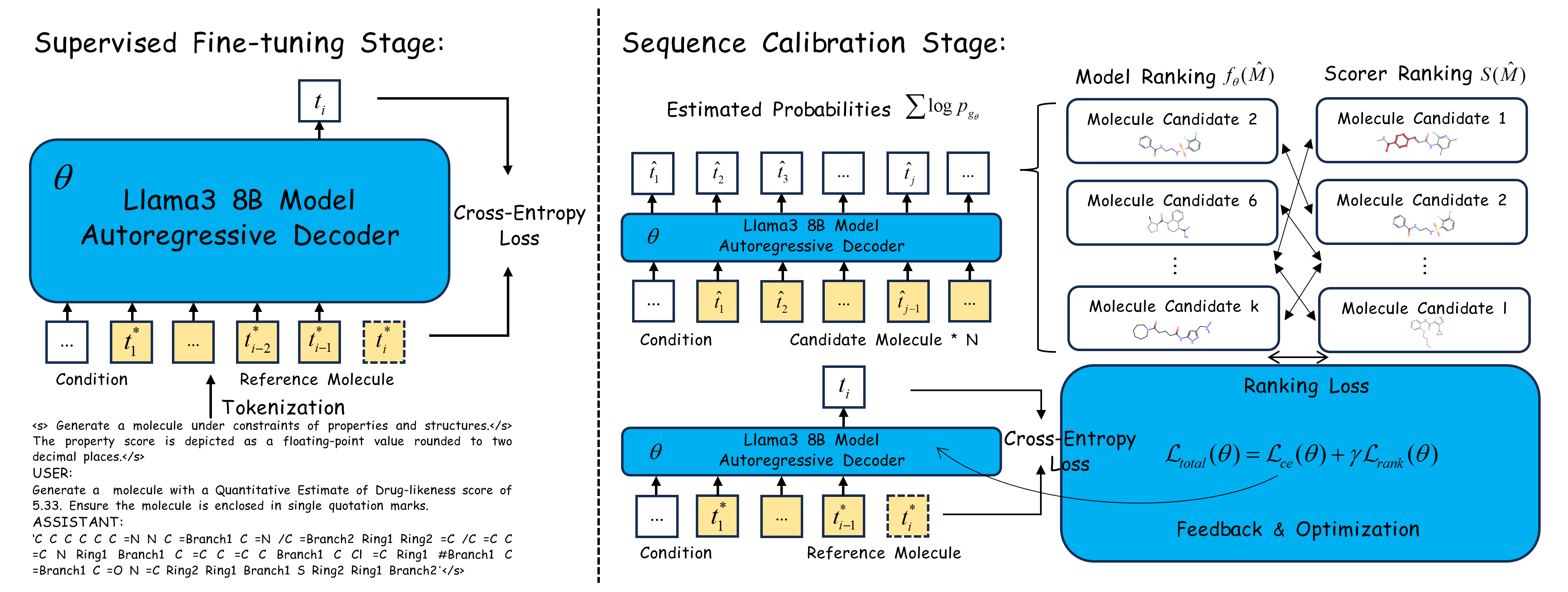}
\caption{Training stages of the ChatMol. The yellow squares below represent tokens used for calculating loss, and the white squares below refer to the condition tokens. During the sequence calibration stage, each training example corresponds to $N$ candidate molecules.}
\label{fig_chatmol}
\end{figure*}

All generation experiments are conducted on eight GeForce RTX 3090 GPUs, and we run the AutoDock-GPU on eight Tesla V100 with 32G memory size. 
We use the parameter-efficient fine-tuning method of LoRA to fine-tune the Llama pre-trained models.
The total learnable parameters are approximately $21$ million, which constitutes about $0.26\%$ of all parameters for Llama3 8B foundation model. 
We employed the bfloat$16$ format for efficiency.
During the supervised fine-tuning stage, ChatMol is trained over $60$ epochs with a learning rate of $2 \times 10^{-5}$ using AdamW optimizer with $0.05$ weight decay, and the same applies to the sequence calibration stage.
We generate $6$ candidates for each training example considering the computational constraints and the two-stage training process was followed as illustrated in \cref{fig_chatmol}.

We generate 100k molecules for single-property logP targeting task, 10k for substructure-property constrained logP extremization task, and 100k molecules for each protein in multi-objective binding affinity maximization task.

\section{Visualization of Generated Molecules}
\begin{figure*}[ht]
\centering
\includegraphics[width=0.85\textwidth]{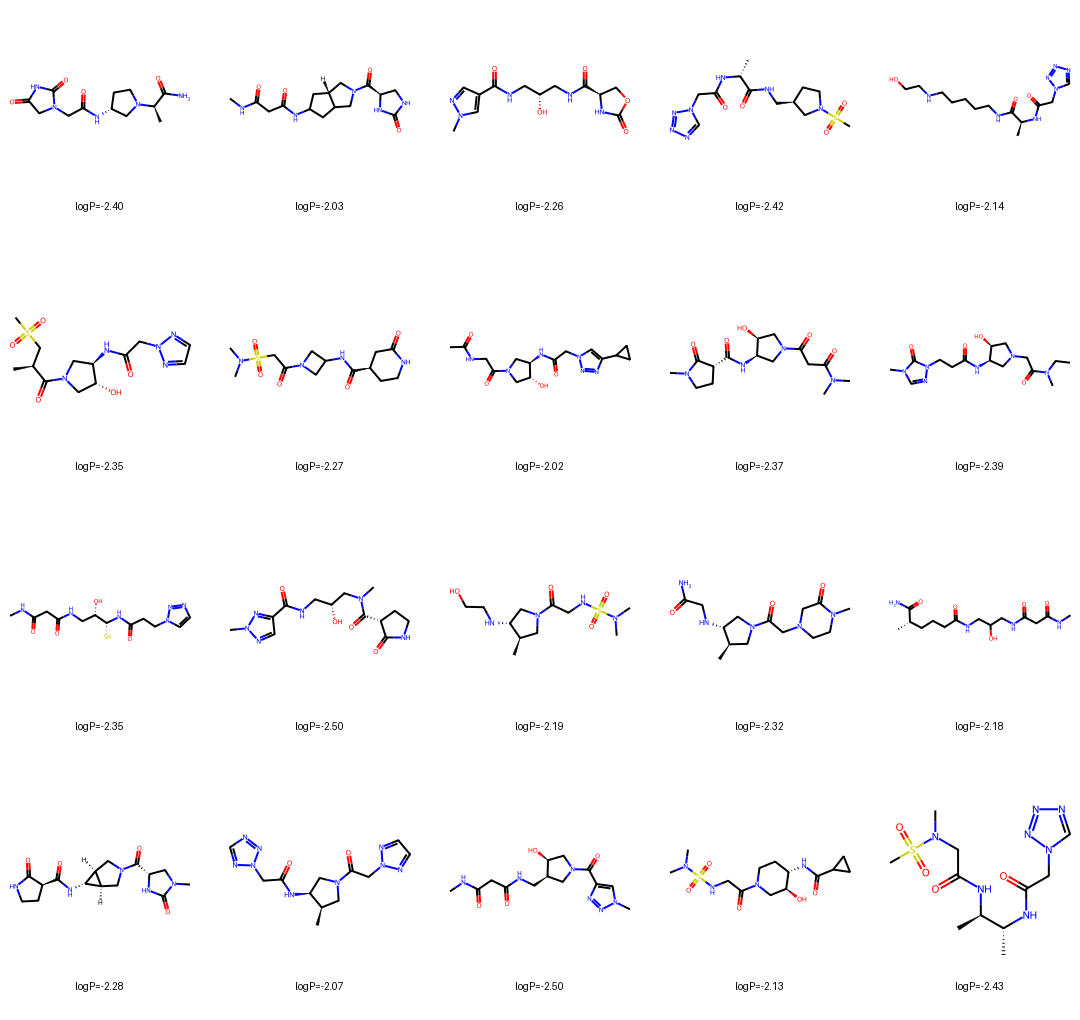}
\caption{Molecules generated in the single-objective logP targeting task.}
\label{fig_target}
\end{figure*}

\begin{figure*}[ht]
\centering
\includegraphics[width=0.85\textwidth]{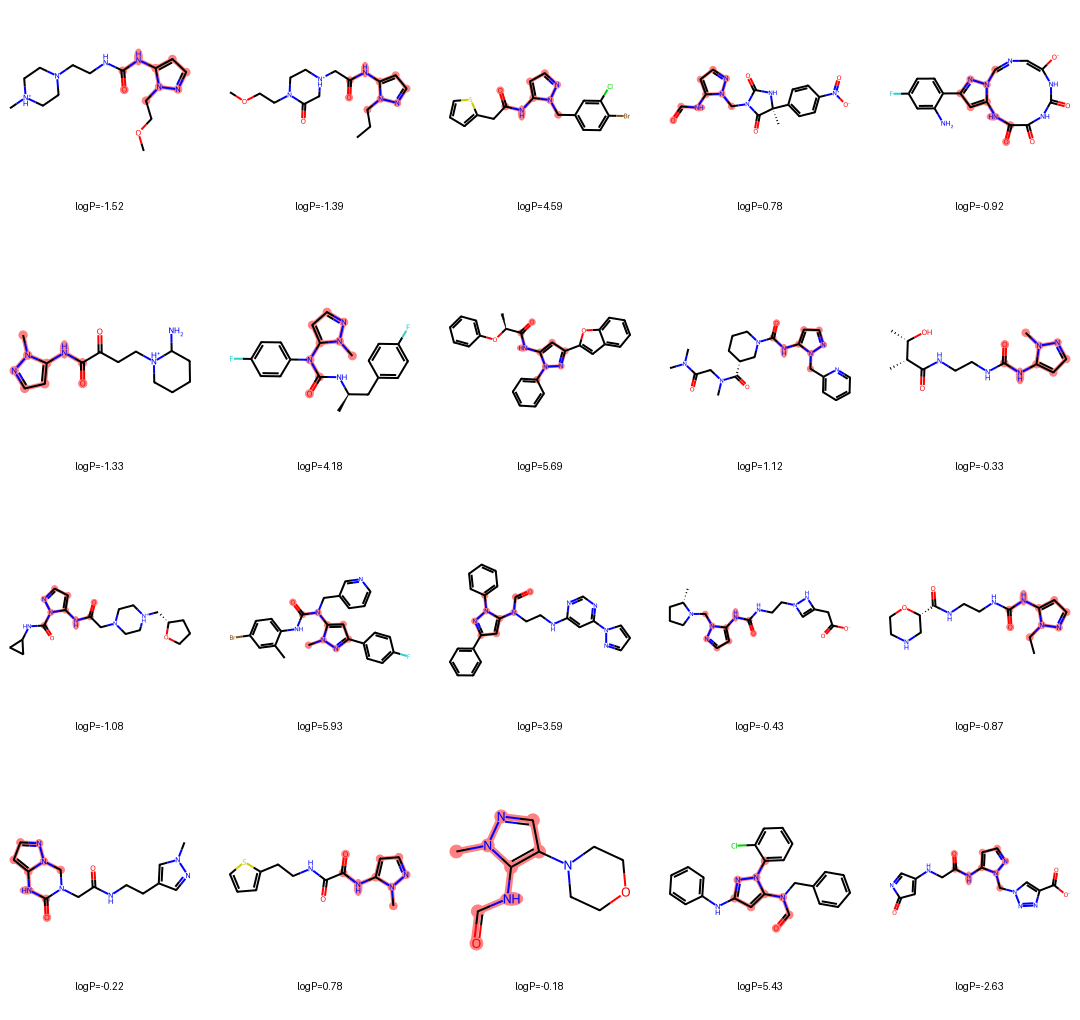}
\caption{Molecules generated in the substructure-property constrained logP extremization task for substructure1. The specified substructures are highlighted in red.}
\label{fig_sub1}
\end{figure*}

\begin{figure*}[ht]
\centering
\includegraphics[width=0.85\textwidth]{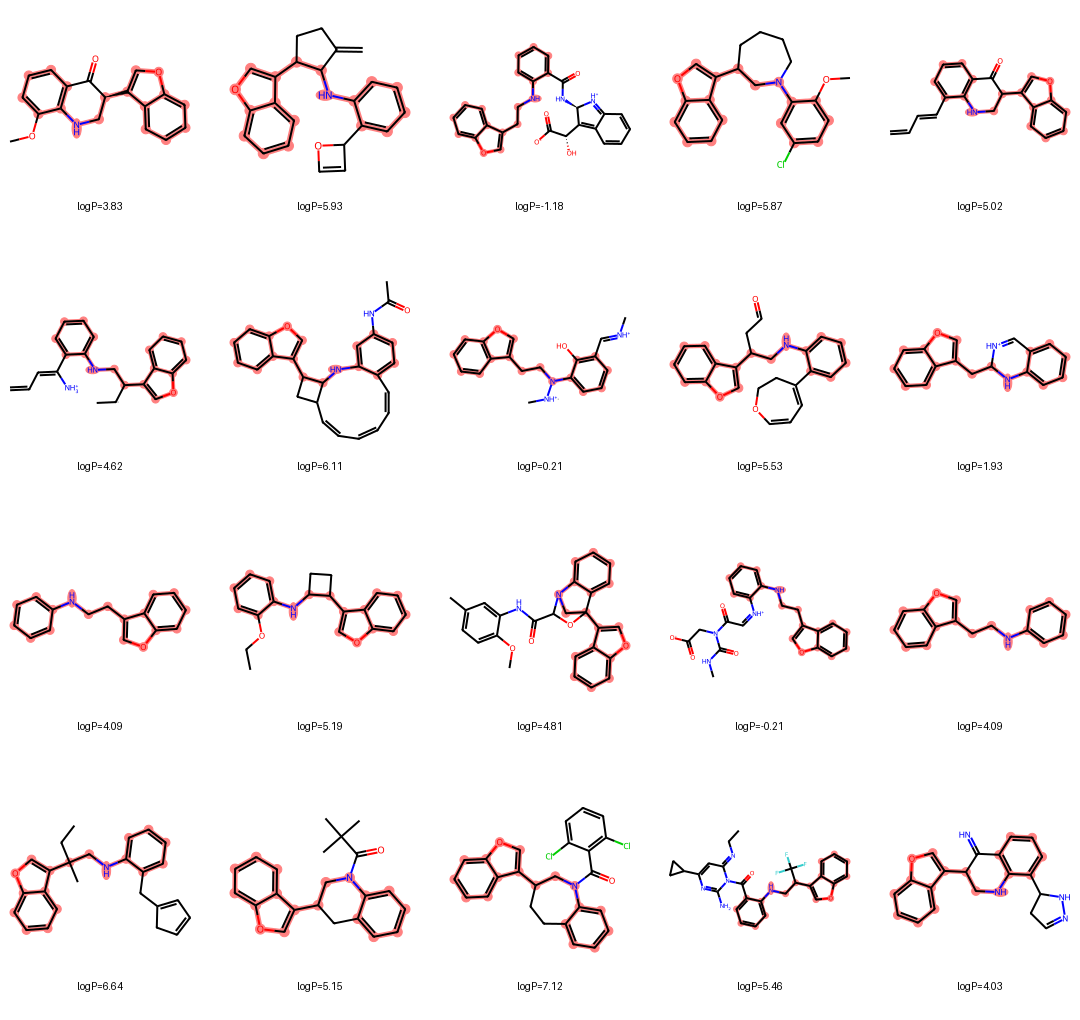}
\caption{Molecules generated in the substructure-property constrained logP extremization task for substructure2. The specified substructures are highlighted in red.}
\label{fig_sub2}
\end{figure*}

\begin{figure*}[ht]
\centering
\includegraphics[width=0.85\textwidth]{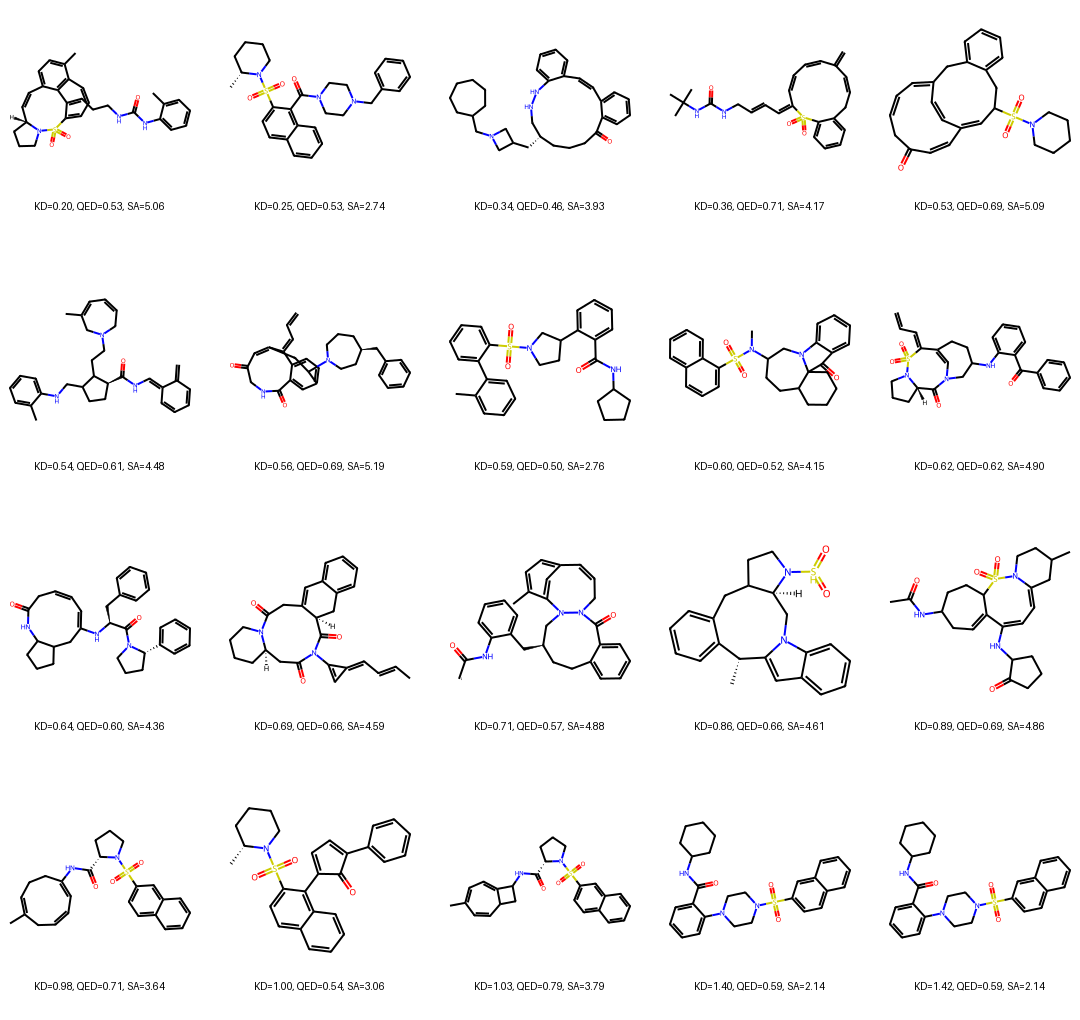}
\caption{Molecules generated in the multi-property constrained binding affinity maximization task for ESR1. The legends denote $\mathrm{K_D} (\downarrow)$, QED $(\uparrow)$, SA $(\downarrow)$.}
\label{fig_1err}
\end{figure*}

\begin{figure*}[ht]
\centering
\includegraphics[width=0.85\textwidth]{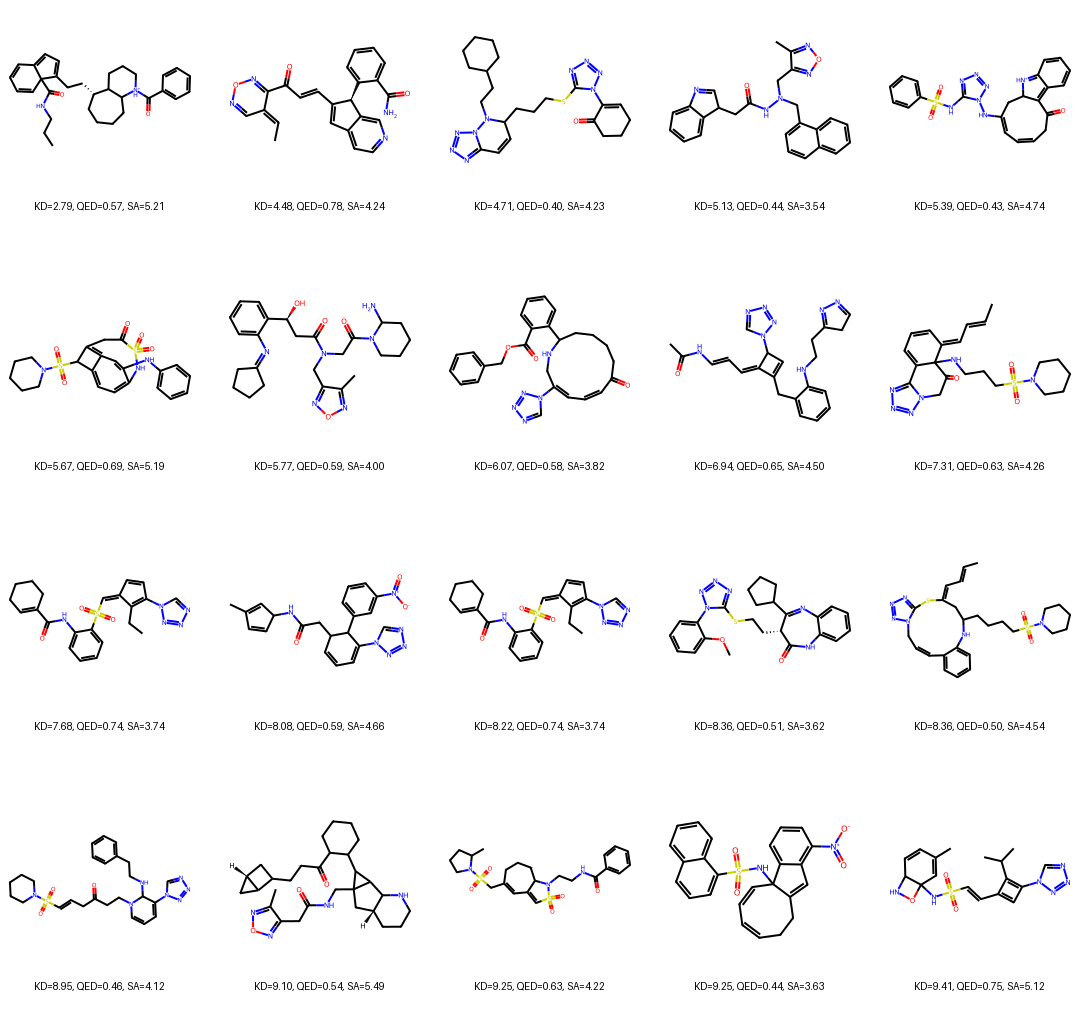}
\caption{Molecules generated in the multi-property constrained binding affinity maximization task for ACAA1. The legends denote $\mathrm{K_D} (\downarrow)$, QED $(\uparrow)$, SA $(\downarrow)$.}
\label{fig_2iik}
\end{figure*}

A subset of generated molecules from the single-property logP targeting task is visualized in \cref{fig_target}. Molecules generated from the substructure-property constrained task are shown in \cref{fig_sub1} and \cref{fig_sub2}, while those from the multi-property constrained task are presented in \cref{fig_1err} and \cref{fig_2iik}.
The specified substructures are highlighted in red.

\section{Limitations}
The numerical values we referred to are those appearing in three downstream tasks, and are not arbitrary.
Their range is limited to the vicinity of the values observed within the tasks.
These numerical embeddings are confined to specific ranges relevant to our tasks.
Since the current values are concentrated at the edges of the molecule property distribution in the training set, which represents a relatively fixed range, our model is minimally affected by the cyclic nature of the position encoding.

RDKit is used as an example scorer, but our framework is indeed flexible and can be extended to incorporate other types of scoring mechanisms.
As long as the scoring system can effectively assess the quality of the generated molecules, it can be used within our framework.
We believe this flexibility allows the model to handle a variety of challenging properties, and we are confident that properties can be addressed by selecting appropriate features for ranking the candidate molecules.

We acknowledge that the training set for multi-property tasks is relatively small. While the model improves compliance, the limited size of relevant training data (particularly in the ZINC250k dataset~\cite{irwin2012zinc}) constrains the model's performance. Nevertheless, we have successfully pushed the boundaries of performance for this task.
\end{document}